\documentclass{ieeeaccess}

\usepackage{graphicx}
\usepackage{latexsym}
\usepackage{amsmath}
\usepackage{amsthm}

\usepackage{amsfonts}
\usepackage{subfigure}
\usepackage{dsfont}
\usepackage{epstopdf}
\usepackage{threeparttable}
\usepackage{multirow}

\usepackage{enumerate}

\usepackage{epsfig}

\usepackage{bm}

\usepackage{cite}
\usepackage{algorithmic}

\usepackage{textcomp}
\usepackage{caption}

\def\BibTeX{{\mathrm B\kern-.05em{\sc i\kern-.025em b}\kern-.08em
    T\kern-.1667em\lower.7ex\hbox{E}\kern-.125emX}}

\begin{document}
\history{Date of publication 0000 00, 0000, date of current version 0000 00, 0000.}
\doi{ }

\title{ELM-based Superimposed CSI Feedback for FDD Massive MIMO System}
\author{\uppercase{Chaojin Qing}\authorrefmark{1}, \IEEEmembership{Member, IEEE},
\uppercase{Bin Cai}\authorrefmark{1}, \uppercase{Qingyao Yang}\authorrefmark{1}, \uppercase{Jiafan Wang}\authorrefmark{2}, \uppercase {and Chuan Huang}\authorrefmark{3},
\IEEEmembership{Member, IEEE}}
\address[1]{School of Electrical Engineering and Electronic Information, Xihua University, Chengdu, 610039, China. (e-mail: qingchj@uestc.edu.cn)}
\address[2]{Synopsys Inc., 2025 NE Cornelius Pass Rd, Hillsboro, OR 97124, USA.}
\address[3]{National Key Laboratory of Science and Technology on Communications, University of Electronic Science and Technology of China, Chengdu, 611731, China.}

\tfootnote{This work is supported in part by the National Key Research and Development Program (Grant 2018YFB1800800), the Key Projects of Education Department of Sichuan Province (Grant 15ZA0134), the Key Scientific Research Fund of Xihua University (Grant Z1120941), and the Major Special Funds of Science and Technology of Sichuan Science and Technology Plan Project (Grant 19ZDZX0016) of China.}

\markboth
{Chaojin Qing \headeretal: ELM-based Superimposed CSI Feedback for FDD Massive MIMO System}
{Chaojin Qing \headeretal: ELM-based Superimposed CSI Feedback for FDD Massive MIMO System}

\corresp{Corresponding author: Chaojin Qing (e-mail: qingchj@uestc.edu.cn).}

\begin{abstract}
In frequency-division duplexing (FDD) massive multiple-input multiple-output (MIMO), deep learning (DL)-based superimposed channel state information (CSI) feedback has presented promising performance. However, it is still facing many challenges, such as the high complexity of parameter tuning, large number of training parameters, and long training time, etc. To overcome these challenges, an extreme learning machine (ELM)-based superimposed CSI feedback is proposed in this paper, in which the downlink CSI is spread and then superimposed on uplink user data sequence (UL-US) to feed back to base station (BS). At BS, an ELM-based network is constructed to recover both downlink CSI and UL-US. In the constructed ELM-based network, we employ the simplified versions of ELM-based subnets to replace the subnets of DL-based superimposed feedback, yielding less training parameters. Besides, the input weights and hidden biases of each ELM-based subnet are loaded from the same matrix by using its full or partial entries, which significantly reduces the memory requirement. With similar or better recovery performances of downlink CSI and UL-US, the proposed ELM-based method has less training parameters, storage space, offline training and online running time than those of DL-based superimposed CSI feedback.

\end{abstract}


\titlepgskip=-15pt
\maketitle

\section{Introduction}
\label{sec:introduction}
\PARstart{T}{he} massive multiple-input multiple-output (MIMO) brings the fifth generation (5G) wireless communication system many advantages in system capacity and link robustness. However, the premise of these advantages is that the accurate downlink channel state information (CSI) can be obtained by base station (BS) \cite{a1}. In time division duplex (TDD) mode, downlink CSI can be estimated from uplink CSI by using channel reciprocity \cite{a2}. For frequency-division duplexing (FDD) mode, the reciprocity-based downlink CSI is not available due to the difference between uplink and downlink frequency bands \cite{a1}, \cite{a3}. Thus, the downlink CSI in FDD massive MIMO system should be estimated by users and fed back to the BS \cite{a3}.

Although the codebook-based CSI feedback method effectively reduces the feedback overhead at user side, the huge number of BS antennas in massive MIMO system substantially results in a tremendous dimension of codebook, which is too large to be applied in practice \cite{a3}. To alleviate this issue, compressive sensing (CS)-based CSI feedback methods have been proposed in \cite{a3}--\cite{a6}, in which the temporal correlation \cite{a3}, sparse enhancement basis \cite{a4}, and spatial correlation \cite{a4}--\cite{a6} of CSI are developed. However, the downlink CSI is approximately sparse for a specific model rather than a general assumption, which may cause practical problems when the hypothesis is not valid \cite{a17}--\cite{a21}.

Recently, deep learning (DL) methods have been successfully applied to physical-layer of wireless communication, e.g., CSI feedback \cite{a17}--\cite{a21}, modulation recognition \cite{a13}--\cite{a15}, information security \cite{a15_1}\cite{a15_2}, etc. For CSI feedback, the DL-based methods outperformed many existing CS schemes in feedback reduction, yet they still occupy significant uplink bandwidth resource. To avoid the occupation of uplink bandwidth resources, the superimposed CSI feedback was proposed in \cite{a16} and further expanded to DL-based approach in \cite{a7}. From \cite{a7}, the DL-based superimposed CSI feedback illuminates that the recovery of downlink CSI is superior than that of superimposed CSI feedback in \cite{a16} with similar UL-US's recovery performance. Even so, the DL-based scheme is still hindered by many disadvantages, such as the high complexity of parameter tuning, large number of training parameters, long training time, etc. This motivates us to develop ELM-based superimposed CSI feedback to improve the DL-based approach in \cite{a7}.

\subsection{RELATED WORKS}

In FDD massive MIMO system, the DL-based CSI feedback methods have been investigated according to feedback reduction (e.g., \cite{a17}--\cite{a21}), and superimposed CSI feedback (e.g., \cite{a16}\cite{a7}), etc.

For feedback reduction, the DL-based CSI feedback developed in \cite{a17}--\cite{a21} could be classified into two categories. The first category is mainly based on a neural network called CsiNet \cite{a17},  which achieved superior performance over various CS-based CSI feedback. Yet, the time correlation, frequency correlation, spatial correlation, feedback delay and feedback errors, etc., were not considered in CsiNet, and thus lead to limited applications. To remedy these defects, a series of improvement methods have emerged in \cite{a17_1}--\cite{a23_1}. In \cite{a17_1}, a CsiNet long short-term memory (CsiNet-LSTM) was proposed by exploiting the time correlation, thereby suiting for practical application in time-varying channels. The recurrent neural network-based CsiNet in \cite{a22} was developed to capture the temporal and frequency correlations of wireless channels. Considering the spatial correlation among antennas, the bidirectional LSTM (Bi-LSTM) and bidirectional convolutional LSTM (Bi-ConvLSTM) were proposed in \cite{a23}. In addition, the feature extractions of CSI according to multiple resolutions and different domains (e.g., spatial and temporal domains) were presented in \cite{a18} and \cite{a23_1}, respectively. Furthermore, the noisy feedback in \cite{a23_2}, feedback errors and feedback delay in \cite{a23_3} were considered to enrich the CSI feedback of CsiNet. These methods in \cite{a17_1}--\cite{a23_1} substantively enhanced the performance of CsiNet in \cite{a17}. Another category of feedback reduction proposed for DL-based CSI feedback takes into account the quantization operation, e.g., \cite{a19}-- \cite{a21}. In \cite{a19}, a bit-level optimized NN, i.e., the joint convolutional residual network (JC-ResNet), was constructed with both CSI compression and quantization. By employing the multiple-rate CS neural network framework, \cite{a20} compressed and quantized the CSI to improve reconstruction accuracy and decrease storage space. The architecture in \cite{a21}, which was composed of convolutional layers followed by quantization and entropy coding blocks, presented promising performance. Although the DL-based CSI feedback in \cite{a17}--\cite{a21} has achieved significant improvement in feedback reduction compared with the CS-based approaches, yet the uplink bandwidth resources are still seriously occupied due to the massive MIMO scenario.

Without any occupation of uplink bandwidth resources, the superimposed CSI feedback was proposed in \cite{a16} and \cite{a7}. In \cite{a16}, the downlink CSI was spread and then superimposed on uplink user data sequences (UL-US) to feed back to BS. Although the occupation of uplink bandwidth was avoided, the recoveries of downlink CSI and UL-US were deteriorated by superposition interference. The interference cancellation has been a research hotspot in the field of wireless communications \cite{a15_4} \cite{a15_5}. To remedy the defect of \cite{a16}, a DL-based superimposed CSI feedback was proposed in \cite{a7}, which consistently improved the estimation of downlink CSI with similar or better UL-US detection performance. But still, the DL-based superimposed CSI feedback is challenged by long training time, complex parameter tuning, large memory requirement, etc.

Unlike the DL-based CSI feedback methods mentioned above, extreme learning machine (ELM) is a single-hidden layer feed-forward neural network, required no gradient back-propagation \cite{a25}. In addition, its input weights and hidden layer biases are randomly generated, and the output weights are analytically calculated by solving the least squares norm problem \cite{a26}. Specially, the ELM network can directly process complex-valued inputs, and can employ complex-valued weights and biases as well \cite{a26}. Therefore, the ELM processed many advantages, e.g., fast learning speed (hundreds of times faster than that of backpropagation algorithm), good generalization performance \cite{a25}--\cite{a15_6}, etc. Inspired by these advantages, an ELM-based superimposed CSI feedback method is proposed in this paper to improve the DL-based superimposed CSI feedback in \cite{a7}.


\begin{figure*}[!hbtp]
\centering
\includegraphics[scale=0.88]{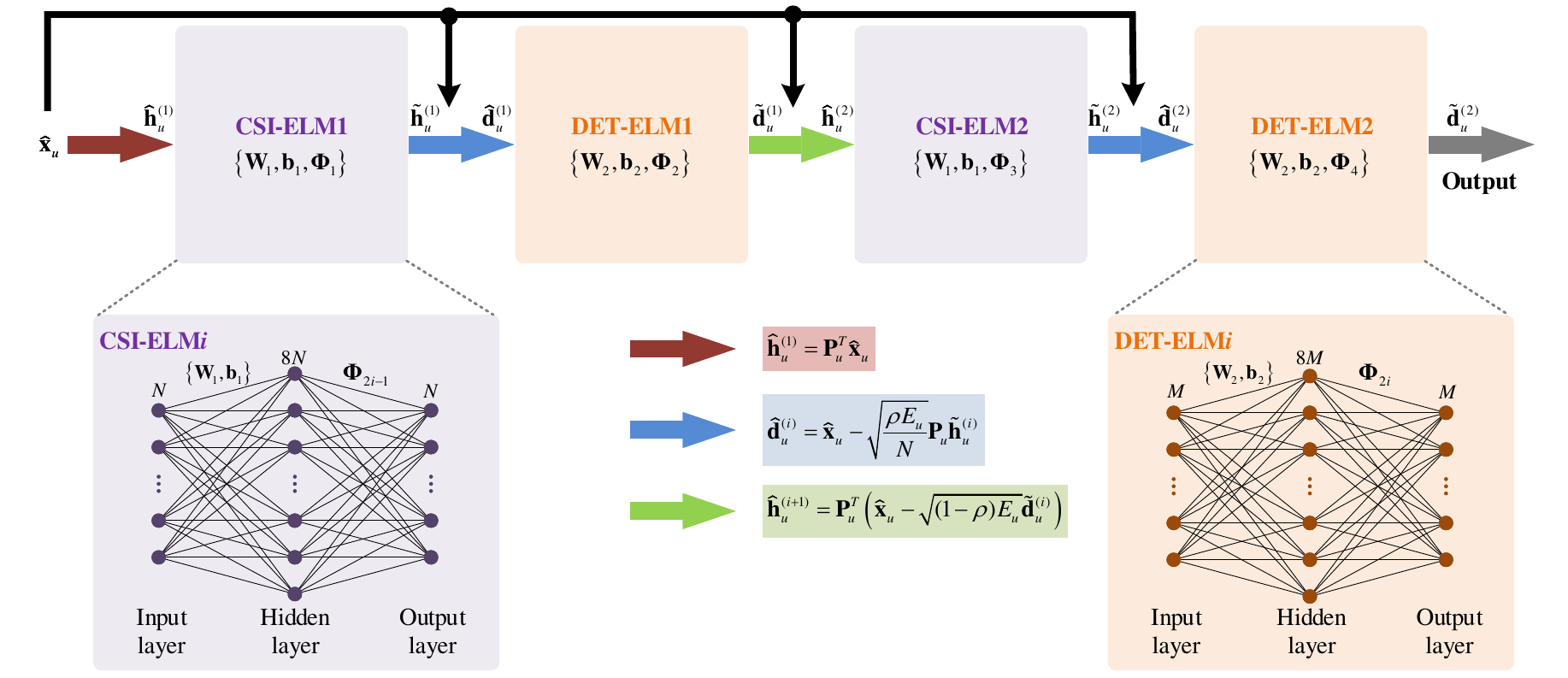}
\caption{Network architecture.}
\label{fig1}
\end{figure*}

\subsection{CONTRIBUTIONS}
 In this paper, an ELM-based superimposed CSI feedback method is proposed to improve the DL-based approach in \cite{a7}. To the best of our knowledge, there are few literatures focusing on the ELM-based superimposed CSI feedback method. The main contributions of this paper are as follows:
\begin{itemize}
\item An ELM-based network is constructed to recover downlink CSI and UL-US, in which four ELM-based subnets are employed to replace and simplify the DL-based subnets in \cite{a7}, leading to low complexity of parameter tuning, less training parameters and training time, etc.

\item In the proposed ELM-based network, the requirement of storage space for input weight matrices and bias vectors of four subnets is greatly reduced, compared with those of the standard ELM network \cite{a28}. This is achieved by loading the input weight matrices and hidden bias vectors of each subnet from the same matrix (which provides full or partial entries).

\item Compared with the DL-based superimposed CSI feedback method in \cite{a7}, the proposed method can obtain a considerable recovery performance of downlink CSI and UL-US with less overhead, e.g., less training parameters, storage space and training time, etc.
\end{itemize}

The remainder of this paper is structured as follows: In Section II, we introduce the system model of superimposed CSI feedback. The ELM-based CSI feedback method is presented in Section III, and followed by the numerical results in Section IV. Finally, Section V concludes our work.

Notations: Bold face upper case and lower case letters denote matrix and vector respectively. ${\left(\cdot \right)^T}$, ${\left(\cdot \right)^H}$, $\mathbf{(\cdot)^\dag}$, denote the transpose, conjugate transpose, and matrix pseudo-inverse respectively. $\mathbf{I}_P$ is the identity matrix of size $P \times P$; ${\mathrm {{\mathbb{BN}}}}\left( \cdot \right)$ denotes the operation of batch normalization; ${\left\|  \cdot  \right\|_2}$ is the Euclidean norm.

\section{SYSTEM MODEL}
Considering a massive MIMO system consists of a BS with $N$ antennas and $U$ single-antenna users, the transmitted signal of user-$u$, $u = 1,2,\ldots,U$, denoted as $\mathbf{x}_u$, can be given by
\begin{equation}\label{EQ1}
{{\mathbf{x}}_{u}}=\sqrt{\frac{\rho {{E}_{u}}}{N}}{{\mathbf{P}}_{u}}{{\mathbf{h}}_{u}}+\sqrt{(1-\rho ){{E}_{u}}}{{\mathbf{d}}_{u}},
\end{equation}
where ${{\rho }}\in \left[ 0,1 \right]$  stands for the power proportional coefficient (PPC) of downlink CSI; $E_u$ represents the transmitted power of user-$u$; ${{\mathbf{h}}_{u}}\in {{\mathbb{C}}^{N\times 1}}$ denotes the downlink CSI from BS to user-$u$; ${{\mathbf{P}}_{u}}\in {{\mathbb{R}}^{M\times N}}$ is a spreading matrix, satisfying $\mathbf{P}_{u}^{T}{{\mathbf{P}}_{u}}=M{{\mathbf{I}}_{N}}$; ${{\mathbf{d}}_{u}}\in {{\mathbb{C}}^{M\times 1}}$ denotes UL-US, and $M$ is the UL-US length. In general, we assume $M>N$ due to main task of the user services.

In (1), the downlink CSI is spread by ${{\mathbf{P}}_{u}}$, and then superimposed on the UL-US to transmit to BS. At BS, after the processing of matched-filter (MF) (i.e., the conventional multiuser detector structure consists of a MF bank front \cite{a27}), the received signal from user-$u$, is denoted as ${{\mathbf{r}}_{u}}\in {{\mathbb{C}}^{N\times M}}$, can be given by \cite{a7}
\begin{equation}\label{EQ2}
{{\mathbf{r}}_{u}}={{\mathbf{g}}_{u}}\mathbf{x}_{u}^{T}+{{\mathbf{n}}_{u}},
\end{equation}
where, ${{\mathbf{n}}_{u}}\in {{\mathbb{C}}^{N\times M}}$ is the additive white Gaussian noise (AWGN) vector of feedback link with zero mean and $\sigma _u^2$-variance entries; ${{\mathbf{g}}_{u}}\in {{\mathbb{C}}^{N\times 1}}$ denotes the uplink channel vector from user-$u$ to BS.

 With the received ${\mathbf{r}}_{u}$ at BS, the main task of superimposed CSI feedback in \cite{a16} and DL-based superimposed CSI feedback in \cite{a7} is to recover downlink CSI and detect UL-US. Compared with the superimposed CSI feedback in \cite{a16}, the DL-based superimposed CSI feedback in \cite{a7} improves the estimation of downlink CSI with similar or better detection performance of UL-US. Even so, the DL-based superimposed CSI feedback in \cite{a7} is still challenging due to long training time, large training parameters, etc. To improve the DL-based superimposed CSI feedback in \cite{a7}, an ELM-based superimposed CSI feedback method is proposed in this paper, which will be elaborated in the next section.

\section{ELM-based superimposed CSI FEEDBACK}
Similar to \cite{a7}, a coarse estimation is also employed by ELM-based superimposed CSI feedback, for which the interference of uplink channel is eliminated and the network structure is simplified. According to the received signal ${{\mathbf{r}}_{u}}$, the coarse estimation can be given by
\begin{equation}\label{EQ3}
{{\mathbf{\widehat{x}}}_{u}^{T}}= \mathbf{g}_{u}^{\dagger }{{\mathbf{r}}_{u}} ={{\mathbf{x}}_{u}^{T}}+ \mathbf{g}_{u}^{\dagger }{{\mathbf{n}}_{u}} .
\end{equation}
Then, the estimated ${{\mathbf{\widehat{x}}}_{u}}$ is delivered to an ELM-based network to recover the downlink CSI and UL-US.

\subsection{NETWORK ARCHITECTURE}
The proposed ELM-based network consists of four subnets (i.e., CSI-ELM1, DET-ELM1, CSI-ELM2 and DET-ELM2), in which the downlink CSI recovery and UL-US detection are addressed by solving a multi-task problem. This network structure is illustrated in Fig.~\ref{fig1} and described as follows:

\begin{itemize}
\item CSI-ELM1 and DET-ELM1 have the same network structure as CSI-ELM2 and DET-ELM2, respectively. CSI-ELM1, DET-ELM1, CSI-ELM2, and DET-ELM2 are successively cascaded to form a multi-task network. Between two cascaded subnets, some expert knowledge is inserted to reduce interference.
\item For CSI-ELM$i$, $i = 1,2$, the neurons of input layer, hidden layer, and output layer are $N$, $8N$, and $N$, respectively. Similarly, the neurons of input layer, hidden layer, and output layer are $M$, $8M$, and $M$ for DET-ELM$i$, respectively. It should be noted that the proposed ELM-based subnets are complex-valued subnet. To match the real-valued subnet in \cite{a7}, its neurons of hidden layer are set as half as that of \cite{a7}.

\item The input of each subnet is normalized by an operation of batch normalization (BN). Different from the standard ELM network \cite{a28}, the hidden output of each subnet employs a linear activation function.
\item Only the output weights (i.e., $\mathbf{\Phi}_1$, $\mathbf{\Phi}_2$, $\mathbf{\Phi}_3$, and $\mathbf{\Phi}_4$) need to be trained. The input weights (i.e., ${{\mathbf{W}}_{1}}$, ${{\mathbf{W}}_{2}}$, ${{\mathbf{W}}_{3}}$ and ${{\mathbf{W}}_{4}}$) and hidden biases (i.e., ${{\mathbf{b}}_{1}}$, ${{\mathbf{b}}_{2}}$, ${{\mathbf{b}}_{3}}$, and ${{\mathbf{b}}_{4}}$) are randomly chosen from the same matrix by using its full or partial entries. Once the input weights and hidden biases are chosen, they are fixed. Compared with the DL-based superimposed CSI feedback network in \cite{a7}, the storage space of parameters, the number of training parameters, and the training time are significantly reduced.

\end{itemize}

As a whole, the proposed ELM-based network possesses the similar architecture as that of the DL-based CSI feedback network in \cite{a7}, but with many advantages, e.g., fewer training parameters, less storage space requirement and shorter training time, etc. These advantages will be presented during the offline training and online running procedures. Naturally, we can further accelerate and simplify neural networks by combining the network compression (e.g., \cite{a27_1}) and ELM. In this paper, we mainly focus on ELM-based superimposed CSI feedback and leave this combination to future work.

\subsection{offline training}
Similar training approach in \cite{a7} (i.e., subnet-by-subnet training) is adopted in this paper, but with many differences. On the one hand, only output weights of each subnet need to be learned for ELM-based network without the requirement of gradient updating \cite{a25}\cite{a28}. On the other hand, the input weights and hidden biases are randomly chosen and then fixed, rather than learned from the ELM-based network. In addition, the way of obtaining the input weights and hidden biases is also different from the standard ELM network \cite{a28}. For the offline training, we first describe the data collection as following.

\subsubsection{DATA COLLECTION}
To train the ELM-based network, the training sets are acquired by a simulation approach. In addition, the input weights and hidden biases also need to be obtained. We first describe the collection of training sets.

In general, only the real-valued data can be processed by DL-based network, e.g., \cite{a17}-\cite{a19},\cite{a7}, etc. Unlike the DL-based network, ELM network can directly process complex-valued data\cite{a26}, avoiding the conversion from complex-valued data to real-valued data. For each subnet, ${{N}_{t}}$ samples are collected for learning its output weights, yielding $4{{N}_{t}}$ samples required for four subnets. Thus, we collect $4{{N}_{t}}$ samples of received signal ${{\mathbf{r}}_{u}}$ and generate input data-sets $\left\{ \left. {{{\mathbf{\widehat{x}}}}_{u,j}} \right|j=1,2,\cdots ,4{{N}_{t}} \right\}$ according to (\ref{EQ3}). Correspondingly, $2{{N}_{t}}$ labels of downlink CSI (i.e., $\left\{ \left. {{{\mathbf{h}}}_{u,j}} \right|j=1,2,\cdots ,2{{N}_{t}}\right\}$) and $2{{N}_{t}}$ labels of UL-US (i.e., $\left\{ \left. {{{\mathbf{d}}}_{u,j}} \right|j=1,2,\cdots ,2{{N}_{t}}\right\}$) are also generated according to (\ref{EQ1}). Then, the training sets for CSI-ELM1, DET-ELM1, CSI-ELM2, and DET-ELM2 can be respectively expressed as
\begin{equation}\label{EQ4}
\left\{ \begin{array}{l}
\left\{ \left( {\mathbf {\widehat{x}}}_{u,1}, {{{\mathbf{h}}}_{u,1}} \right ), \cdots, \left( {\mathbf {\widehat{x}}}_{u,N_t}, {{{\mathbf{h}}}_{u,N_t}} \right )   \right \},\\
\left\{ \left( {\mathbf {\widehat{x}}}_{u,{N_t}+1}, {{{\mathbf{d}}}_{u,1}} \right ), \cdots, \left( {\mathbf {\widehat{x}}}_{u,2N_t}, {{{\mathbf{d}}}_{u,N_t}} \right )   \right \},\\
\left\{ \left( {\mathbf {\widehat{x}}}_{u,{2N_t}+1}, {{{\mathbf{h}}}_{u,{N_t}+1}} \right ), \cdots, \left( {\mathbf {\widehat{x}}}_{u,3N_t}, {{{\mathbf{h}}}_{u,2N_t}} \right )   \right \},\\
\left\{ \left( {\mathbf {\widehat{x}}}_{u,{3N_t}+1}, {{{\mathbf{d}}}_{u,{N_t}+1}} \right ), \cdots, \left( {\mathbf {\widehat{x}}}_{u,4N_t}, {{{\mathbf{d}}}_{u,2N_t}} \right )   \right \}.\\
\end{array} \right.
\end{equation}
For expression convenience, we denote the training sets for CSI-ELM1, DET-ELM1, CSI-ELM2, and DET-ELM2 as $\left ({\mathbf{x}}_u^{(1)}, {\mathbf{T}}_{\mathbf{h}}^{\left( 1 \right)} \right ) $, $\left ({\mathbf{x}}_u^{(2)}, {\mathbf{T}}_{\mathbf{d}}^{\left( 1 \right)} \right) $, $\left ({\mathbf{x}}_u^{(3)}, {\mathbf{T}}_{\mathbf{h}}^{\left( 2 \right)} \right) $, and $ \left ({\mathbf{x}}_u^{(4)}, {\mathbf{T}}_{\mathbf{d}}^{\left( 2 \right)} \right) $, respectively. ${\mathbf{x}}_u^{(k)}$, $\mathbf{T}_{\mathbf{h}}^{\left( i \right)}$, and $\mathbf{T}_{\mathbf{d}}^{\left( i \right)}$ are defined as
\begin{equation}\label{EQ5}
\left\{ \begin{array}{l}
{\mathbf{x}}_u^{(k)} = \left[ {{{\mathbf{\widehat{x}}}_{u,(k - 1){N_t} + 1}}, \cdots ,{{\mathbf{\widehat{x}}}_{u,k{N_t}}}} \right]\\
{\mathbf{T}}_{\mathbf{h}}^{\left( i \right)} = \left[ {{{\mathbf{h}}_{u,\left( {i - 1} \right){N_t} + 1}}, \cdots ,{{\mathbf{h}}_{u, {i} {N_t}}}} \right]\\
{\mathbf{T}}_{\mathbf{d}}^{\left( i \right)} = \left[ {{{\mathbf{d}}_{u,\left( {i - 1} \right){N_t} + 1}}, \cdots ,{{\mathbf{d}}_{u,i{N_t}}}} \right]
\end{array} \right.,
\end{equation}
where $k = 1,2,3,4$ and $ i = 1,2$.

The generation of input weights and hidden biases is different from the standard ELM network \cite{a28}. We randomly generate $\mathbf{W} \in \mathbb{R}^{8M\times M}$, and save it in storage space. In $\mathbf{W}$, each entry is modeled as standard normal distribution with zero mean and unit variance. Then all input weights and hidden biases (i.e., $\mathbf{W}_1$, $\mathbf{W}_2$, $\mathbf{b}_1$, and $\mathbf{b}_2$) are loaded from $\mathbf{W}$ by using its full or partial entries. Since only $8M^2$ coefficients (from $\mathbf{W}$) need to be saved rather than all input weights and hidden biases of standard ELM network, the storage space can be significantly reduced.

\subsubsection{NETWORK TRAINING}
With the collected data, CSI-ELM1, DET-ELM1, CSI-ELM2 and DET-ELM2 are trained in turn to obtain $\mathbf{\Phi}_1$, $\mathbf{\Phi}_2$, $\mathbf{\Phi}_3$, and $\mathbf{\Phi}_4$, respectively. Prior to a subset training, the expert knowledge is employed to eliminate superimposed interference. After a subnet is trained, its input weights, hidden biases and output weights are then fixed for training the following subnet. By referencing the architecture in Fig.~\ref{fig1}, the training procedure is presented as follows.

\textbf{Despreading}: Before training each subnet, a despreading operation for the training set ${\mathbf{x}}_u^{(k)}$, $k=1,2,3,4$, is employed to reduce superimposed interference from UL-US, which can be given by
\begin{equation}\label{EQ6}
{\mathbf{\widehat h}}_u^{(1)} = {\mathbf{P}}_u^T{\mathbf{ x}}_u^{(k)},
\end{equation}
where ${\mathbf{ x}}_u^{(1)}$, ${\mathbf{ x}}_u^{(2)}$, ${\mathbf{ x}}_u^{(3)}$, and ${\mathbf{ x}}_u^{(4)}$, are used to generate CSI-ELM1's input for training CSI-ELM1, DET-ELM1, CSI-ELM2, and DET-ELM2, respectively.

\textbf{Training CSI-ELM$i$}: With the training input ${\mathbf{\widehat h}}_u^{(i)}$, $i=1,2$, the hidden output of CSI-ELM$i$, denoted as ${\mathbf{H}}_{u}^{(2i-1)}$, can be expressed as
\begin{equation}\label{EQ7}
{\mathbf{H}}_{u}^{(2i-1)}{\rm{ = }}{{\mathbf{W}}_1}\left( {{{\mathbb{BN}}}\left( {{\mathbf{\widehat h}}_{u}^{(i)}} \right)} \right) + {{\mathbf{b}}_1}.
\end{equation}
In (\ref{EQ7}), ${\mathbf{\widehat h}}_u^{(1)}$ is obtained according to (\ref{EQ6}), while the ${\mathbf{\widehat h}}_u^{(2)}$ is formed based on the output of DET-ELM1. According to the hidden output matrix ${\mathbf{H}}_u^{(2i-1)}$ and label $\mathbf{T}_{\mathbf{h}}^{\left( i \right)}$, the output weight matrix of CSI-EML$i$ can be given by
\begin{equation}\label{EQ8}
{{\mathbf{\Phi }}_{2i-1}} = {\mathbf{T}}_{\mathbf{h}}^{\left( i \right)}{\left( {{\mathbf{H}}_u^{(2i-1)}} \right)^\dag }.
\end{equation}
Once the output weight matrix ${{\mathbf{\Phi }}_{2i-1}}$ was obtained, it was then fixed for training the following subnet.

\textbf{Reduction of downlink CSI interference}: In order to train DET-ELM$i,i=1,2$, we use the expert knowledge to reduce the superimposed interference from downlink CSI. According to the output of CSI-ELM$i$, i.e., ${\mathbf{\widetilde h}}_u^{(i)}$, this interference reduction can be represented as
\begin{equation}\label{EQ9}
{\mathbf{\widehat d}}_u^{(i)} = {\mathbf{\widehat x}}_u^{(2i)} - \sqrt {\frac{{\rho {E_u}}}{N}} {{\mathbf{P}}_u}{\mathbf{\widetilde h}}_u^{(i)}.
\end{equation}
From (\ref{EQ9}), the superimposed interference from downlink CSI is partly removed, yielding an improved input of DET-ELM$i$. With the input ${\mathbf{\widehat d}}_u^{(i)}$, DET-ELM$i$ is then trained to learn its output weight ${{\mathbf{\Phi }}_{2i}}$.

\textbf{Training DET-ELM$i$}: According to the input ${\mathbf{\widehat d}}_u^{(i)}$, $i=1,2$, the hidden output of DET-ELM$i$, denoted as ${\mathbf{H}}_{u}^{(2i)}$, can be given by
\begin{equation}\label{EQ10}
{\mathbf{H}}_{u}^{(2i)}{\rm{ = }}{{\mathbf{W}}_2}\left( {{\mathbb{BN}}\left( {{\mathbf{\widehat d}}_{u}^{(i)}} \right)} \right) + {{\mathbf{b}}_2}.
\end{equation}
Based on the hidden output ${\mathbf{H}}_{u}^{(2i)}$ and label $\mathbf{T}_{\mathbf{d}}^{\left( i \right)}$, the output weight matrix of DET-EML$i$ can be expressed as
\begin{equation}\label{EQ11}
{{\mathbf{\Phi }}_{2i}} = {\mathbf{T}}_{\mathbf{d}}^{\left( i \right)}{\left( {{\mathbf{H}}_u^{(2i)}} \right)^\dag }.
\end{equation}
After the training of DET-ELM$i$ is finished, the output weight matrix ${{\mathbf{\Phi }}_{2i}}$ is obtained. Then, the learned ${{\mathbf{\Phi }}_{2}}$ (i.e., $i=1$) is used to train the next subnet (i.e., CSI-ELM2) and the learned ${{\mathbf{\Phi }}_{4}}$ (i.e., $i=2$) is saved for online running.

\textbf{UL-US interference reduction}: With the trained ${{\mathbf{\Phi }}_{2}}$, the DET-ELM$1$ can produce its output ${\mathbf{\widetilde d}}_u^{(1)}$. Before entering CSI-ELM$2$, the superimposed interference from UL-US should be reduced. This interference reduction can be represented as
\begin{equation}\label{EQ12}
{\mathbf{\widehat h}}_u^{(2)} = {\mathbf{P}}_u^T\left({\mathbf{\widehat x}}_u^{(3)} - \sqrt {(1 - \rho ){E_u}} {\mathbf{\widetilde d}}_u^{(1)}\right).
\end{equation}
Then, the superimposed interference from UL-US is partly removed, which improves the input of CSI-ELM$2$.

From the processing mentioned above, the output weights, i.e., $\mathbf{\Phi}_1$, $\mathbf{\Phi}_2$, $\mathbf{\Phi}_3$, and $\mathbf{\Phi}_4$, are learned from the network training. This training procedure is summarized in TABLE~\ref{table_I}.

\begin{table}[!ht]
\renewcommand\arraystretch{1.2}
\caption{TRAINING PROCEDURE}
\label{table_I}
\begin{tabular}{l}
\hline
\hline

\kern -2pt \textbf{Input:} The training sets for CSI-ELM1, DET-ELM1, CSI-ELM2, and \\
\kern 22pt DET-ELM2. \\

\hline
\kern 4pt 1): Train CSI-ELM1 to obtain output weight matrix ${{\mathbf{\Phi }}_{1}}$ according to \\
    \kern 15pt training sets $\left ({\mathbf{x}}_u^{(1)}, {\mathbf{T}}_{\mathbf{h}}^{\left( 1 \right)} \right ) $.\\

\kern 4pt 2): Keeping the parameters of CSI-ELM1 unchanged, based on training \\
\kern 15pt  sets $\left ({\mathbf{x}}_u^{(2)}, {\mathbf{T}}_{\mathbf{d}}^{\left( 1 \right)} \right ) $, train DET-ELM1 to achieve ${{\mathbf{\Phi }}_{2}}$.\\

\kern 4pt 3): Maintaining CSI-ELM1 and DET-ELM1's parameters unchanged, \\
    \kern 15pt train CSI-ELM2 to get ${{\mathbf{\Phi }}_{3}}$ by training sets $\left ({\mathbf{x}}_u^{(3)}, {\mathbf{T}}_{\mathbf{h}}^{\left( 2 \right)} \right ) $.\\

\kern 4pt 4): Remaining CSI-ELM1, DET-ELM1 and CSI-ELM2's parameters \\
    \kern 15pt unchanged, train DET-ELM2 to obtain ${{\mathbf{\Phi }}_{4}}$ according to training\\
    \kern 15pt sets $\left ({\mathbf{x}}_u^{(4)}, {\mathbf{T}}_{\mathbf{d}}^{\left( 2 \right)} \right ) $.\\
\hline

\kern -2pt \textbf{Output and save:} ${{\mathbf{\Phi }}_1},{{\mathbf{\Phi }}_2},{{\mathbf{\Phi }}_3}$ and ${{\mathbf{\Phi }}_4}$.\\

 \hline
 \hline
\end{tabular}
\end{table}

According to the network training and corresponding processing, all required network parameters, i.e., $\left\{ {{{\mathbf{W}}_1},{{\mathbf{W}}_2},{{\mathbf{b}}_1},{{\mathbf{b}}_2},{{\mathbf{\Phi }}_1},{{\mathbf{\Phi }}_2},{{\mathbf{\Phi }}_3},{{\mathbf{\Phi }}_4}} \right\}$, have been determined. Relative to the training of DL-based network in \cite{a7}, the training for proposed ELM-based network presents many advantages. On the one hand, the gradient back-propagation is usually employed by DL-based network's training to update network parameters and minimize the loss function \cite{a28}. This training usually needs large training set and long training time. On the other hand, the DL-based network's training is usually accompanied by gradient disappearance, over-fitting, and complex parameter tuning \cite{a25}\cite{a28_1}, etc. In contrast, the proposed ELM-based network is a forward network (without the requirement of gradient back-propagation), whose parameter training can be performed by matrix operation \cite{a28}. Since only the output weight matrices (i.e., $\mathbf{\Phi}_1$, $\mathbf{\Phi}_2$, $\mathbf{\Phi}_3$, and $\mathbf{\Phi}_4$) rather than all network parameters (i.e., $ {{{\mathbf{W}}_1},{{\mathbf{W}}_2},{{\mathbf{b}}_1},{{\mathbf{b}}_2},{{\mathbf{\Phi }}_1},{{\mathbf{\Phi }}_2},{{\mathbf{\Phi }}_3},{{\mathbf{\Phi }}_4}} $) need to be trained, the proposed ELM-based network possesses shorter training time, less training parameters, and easier training operation than DL-based network in \cite{a7}. In addition, ${{\mathbf{W}}_1}$, ${{\mathbf{W}}_2}$, ${{\mathbf{b}}_1}$, and ${{\mathbf{b}}_2}$ are all loaded from the same matrix ${{\mathbf{W}}}$ (i.e., the actual memory for all these parameters is equal to the size of matrix ${{\mathbf{W}}})$, leading to less parameter memory compared with that of \cite{a7}.

\subsection{online running}

With the trained network parameters, the online running procedure is presented in TABLE~\ref{table_II}, and some explications are given as follows.

\textbf{Network input}: With the received signal ${{\mathbf{r}}_{u}}$ in (\ref{EQ2}), the coarse estimation is employed according to (\ref{EQ3}) to capture the network input ${{\mathbf{\widehat x}}_u} \in {{\mathbb{C}}^{M \times 1}}$ and simplify network architecture.

\textbf{Estimation of downlink CSI}: The CSI-ELM1 and CSI-ELM2 are employed to estimate downlink CSI, which are given in step 3) and step 7) of TABLE~\ref{table_II}, respectively. This estimation of CSI-ELM$i$, $i=1,2$, can be expressed as
\begin{equation}\label{EQ13}
{\mathbf{\widetilde h}}_u^{(i)}{\rm{ = }}{{\mathbf{\Phi }}_{2i - 1}}\left\{ {{{\mathbf{W}}_1}\left( {{\mathbb{BN}}\left( {{\mathbf{\widehat h}}_u^{(i)}} \right)} \right) + {{\mathbf{b}}_1}} \right\},
\end{equation}
where ${{\mathbf{\widehat h}}_u^{(1)}}$ and ${{\mathbf{\widehat h}}_u^{(2)}}$ are the outputs of despreading and interference reduction (from UL-US), respectively. In TABLE~\ref{table_II}, ${{\mathbf{\widehat h}}_u^{(1)}}$ and ${{\mathbf{\widehat h}}_u^{(2)}}$ can be obtained according to step 2) and step 6), respectively. Specially, ${\mathbf{\widetilde h}}_u^{(2)}$ is viewed as the estimated downlink CSI according to the proposed ELM-based network in this paper.

\textbf{Reduction of downlink CSI interference}: Prior to UL-US's detection, the interferences from downlink CSI are reduced according to the step 4) and step 8) in TABLE 2 for DET-ELM1 and DET-ELM2, respectively. The interference reduction for DET-ELM$i$, $i=1,2$, can be represented as
\begin{equation}\label{EQ14}
{\mathbf{\widehat d}}_u^{(i)} = {\mathbf{\widehat x}}_u - \sqrt {\frac{{\rho {E_u}}}{N}} {{\mathbf{P}}_u}{\mathbf{\widetilde h}}_u^{(i)}.
\end{equation}
Then, ${\mathbf{\widehat d}}_u^{(1)}$ and ${\mathbf{\widehat d}}_u^{(2)}$ are used to serve the UL-US's detections of DET-ELM1 and DET-ELM2, respectively.


\begin{table}[!ht]
\renewcommand\arraystretch{1.2}
\caption{ONLINE RUNNING PROCEDURE}
\label{table_II}
\begin{tabular}{l}
\hline
\hline

\textbf{Input:} The received signal from user-$u$ at BS, i.e., ${{\mathbf{r}}_{u}}$.\\

\hline
\kern 4pt 1): Perform a coarse estimation to obtain ${{\mathbf{\widehat x}}_u}$, according to (\ref{EQ3}).\\

\kern 4pt 2): Despread ${{\mathbf{\widehat x}}_u}$ according to (\ref{EQ6}), i.e., ${\mathbf{\widehat h}}_u^{(1)} = {\mathbf{P}}_u^T{\mathbf{\widehat x}}_u$.\\

\kern 4pt 3): Use CSI-ELM1 to estimate downlink CSI with input ${\mathbf{\widehat h}}_u^{(1)}$, which is\\
    \kern 15pt expressed in (\ref{EQ13}) with $i=1$, and obtain the estimation ${\mathbf{\widetilde h}}_u^{(1)}$.\\

\kern 4pt 4): Reduce downlink CSI interference according to (\ref{EQ14}) with $i=1$,\\
    \kern 15pt i.e., use the expert knowledge to obtain ${\mathbf{\widehat d}}_u^{(1)}$ according to ${\mathbf{\widetilde h}}_u^{(1)}$.\\

\kern 4pt 5): Utilize DET-ELM1 to detect UL-US with input ${{\mathbf{\widehat d}}_u^{(1)}}$, i.e., perform \\
    \kern 15pt a detection to obtain ${\mathbf{\widetilde d}}_u^{(1)}$, which is expressed in (\ref{EQ15}) with $i=1$.\\

\kern 4pt 6): Eliminate UL-US interference according to (\ref{EQ16}), i.e., use the expert \\
    \kern 15pt knowledge to obtain ${\mathbf{\widehat h}}_u^{(2)}$ according to ${\mathbf{\widetilde d}}_u^{(1)}$.\\

\kern 4pt 7): Use CSI-ELM2 to estimate downlink CSI with input ${{\mathbf{\widehat h}}_u^{(2)}}$, which is \\
    \kern 15pt  represented in (\ref{EQ13}) with $i=2$, where the estimated downlink CSI is\\
     \kern 15pt denoted by ${\mathbf{\widetilde h}}_u^{(2)}$.\\

\kern 4pt 8): Remove downlink CSI interference according to (\ref{EQ14}) with $i=2$, \\
\kern 15pt i.e., use the expert knowledge to obtain ${\mathbf{\widehat d}}_u^{(2)}$.\\

\kern 4pt 9): Utilize DET-ELM2 to detect UL-US with input ${{\mathbf{\widehat d}}_u^{(2)}}$, i.e., detect to\\
 \kern 15pt  obtain ${\mathbf{\widetilde d}}_u^{(2)}$, which is expressed in (\ref{EQ15}) with $i=2$.\\

\hline

\textbf{Output:} ${{\mathbf{\widetilde h}}_u} = {\mathbf{\widetilde h}}_u^{\left( 2 \right)}$ and ${{\mathbf{\widetilde d}}_u} = {\mathbf{\widetilde d}}_u^{\left( 2 \right)}$.\\

 \hline
 \hline
\end{tabular}
\end{table}

\textbf{Detection of UL-US}: Based on the captured ${\mathbf{\widehat d}}_u^{(i)}$, $i=1,2$, the UL-US can be detected by using DET-ELM$i$, which can be expressed as
\begin{equation}\label{EQ15}
{\mathbf{\widetilde d}}_u^{(i)}{\rm{ = }}{{\mathbf{\Phi }}_{2i}}\left\{ {{{\mathbf{W}}_2}\left( {{\mathbb{BN}}\left( {{\mathbf{\widehat d}}_u^{(i)}} \right)} \right) + {{\mathbf{b}}_2}} \right\}.
\end{equation}
In (\ref{EQ15}), ${\mathbf{\widetilde d}}_u^{(1)}$ and ${\mathbf{\widetilde d}}_u^{(2)}$ are detected according to step 5) and step 9) in TABLE~\ref{table_II}, respectively. In the proposed ELM-based network, ${\mathbf{\widetilde d}}_u^{(2)}$ is used as the ultimate UL-US's detection, while the ${\mathbf{\widetilde d}}_u^{(1)}$ is employed as the intermediate variable for generating the input of ELM-ELM2.

\textbf{UL-US interference reduction}: With the detected ${\mathbf{\widetilde d}}_u^{(1)}$, we employ an interference reduction, i.e., the step 6) in TABLE~\ref{table_II}, to generate the ELM-ELM2's input, which can be given by
\begin{equation}\label{EQ16}
{\mathbf{\widehat h}}_u^{(2)} = {\mathbf{P}}_u^T\left({{\mathbf{\widehat x}}_u} - \sqrt {(1 - \rho ){E_u}} {\mathbf{\widetilde d}}_u^{(1)}\right).
\end{equation}
Then, ${\mathbf{\widehat h}}_u^{(2)}$ is used as the input of CSI-ELM2 to estimate downlink CSI, which is given in step 7) in TABLE~\ref{table_II}.

From step 1) to step 9) in TABLE~\ref{table_II}, the downlink CSI and UL-US can be recovered according to the online running, i.e., ${{\mathbf{\widetilde h}}_u} = {\mathbf{\widetilde h}}_u^{\left( 2 \right)}$ and ${{\mathbf{\widetilde d}}_u} = {\mathbf{\widetilde d}}_u^{\left( 2 \right)}$ can be obtained from the proposed ELM-based network with the received signal ${\mathbf{r}}_u$.

\section{EXPERIMENTAL ANALYSIS}
In this section, we give some numerical results of the proposed ELM-based CSI feedback. Some definitions and basic parameters involved in simulations are first given in IV-A. Then, we show the normalized mean squared error (NMSE) of downlink CSI and bit error rate (BER) of UL-US in IV-B to verify the effectiveness of the proposed ELM-based CSI feedback. The last but not the least, in IV-C, the less training parameters, storage space, training time and online running time than those of DL-based superimposed feedback in \cite{a7} are presented. In the following, we describe the experimental setting.

\subsection{parameter setting}
Some definitions involved in simulations are given as follows. The signal-to-noise ratio (SNR) in decibel (dB) of the signal received at BS from user--$u$ is defined as \cite{a16}\cite{a7}
\begin{equation}\label{EQ17}
SNR = 10{\log _{10}}\left( {\frac{{{E_u}}}{{\sigma _u^2}}} \right).
\end{equation}
The NMSE, which is used to evaluate the recovery of downlink CSI, is defined as \cite{a16}\cite{a7}
\begin{equation}\label{EQ18}
NMSE = {\rm{E}}\left( {\frac{{\left\| {{{{\mathbf{\widetilde h}}}_u} - {{\mathbf{h}}_u}} \right\|_2^2}}{{\left\| {{{\mathbf{h}}_u}} \right\|_2^2}}} \right).
\end{equation}
In the experiment phase, $M = 512$, ${N_t} = 10^4$. The Walsh matrix is employed as the spreading matrix ${{\mathbf{P}}_u}$. Both uplink and downlink CSI, i.e., ${{\mathbf{h}}_u}$ and ${{\mathbf{g}}_u}$, are randomly generated on the basis of the distribution $\mathcal{CN}\left( {0,\left( {{1 \mathord{\left/{\vphantom {1 N}} \right. \kern-\nulldelimiterspace} N}} \right)} \right)$. The UL-US ${\mathbf{d}}_u$ is formed according to the symbols of quadrature-phase-shift-keying (QPSK) modulation. We randomly generate the matrix ${{\mathbf{W}}}$, i.e., each entry in ${{\mathbf{W}}}$ obeys the independent and identically distributed (i.i.d.) Gaussian distribution with zero mean and unit variance. The training input data-sets are generated from (\ref{EQ1}) to (\ref{EQ3}), where $\rho=0.20$ is considered. Unlike DL-based network in \cite{a7}, where the training SNR is set as $SNR=5$dB, the ELM-based network is trained under noise-free case. According to \cite{a7}, the hidden neurons of the CSI-NET$i$ and the DET-NET$i$, which are employed in DL-based superimposed feedback, are $16N$ and $16M$, respectively. The testing data is generated by utilizing the same method of generating the training data, and we stop the testing for BER performance when at least 1000-bit errors are observed \cite{a7}.

For description convenience in IV-B and IV-C, we denote the traditional superimposed feedback in \cite{a16}, the DL-based superimposed feedback in \cite{a7}, and the proposed ELM-based superimposed CSI feedback as ``Ref\cite{a16}'', ``Ref\cite{a7}'', and ``Proposed'', respectively.

\begin{figure*}[!hbtp]
\centering
    \subfigure[$N=16$]{
    \label{figure2_support_set_ber}
    \includegraphics[width=0.63\columnwidth]{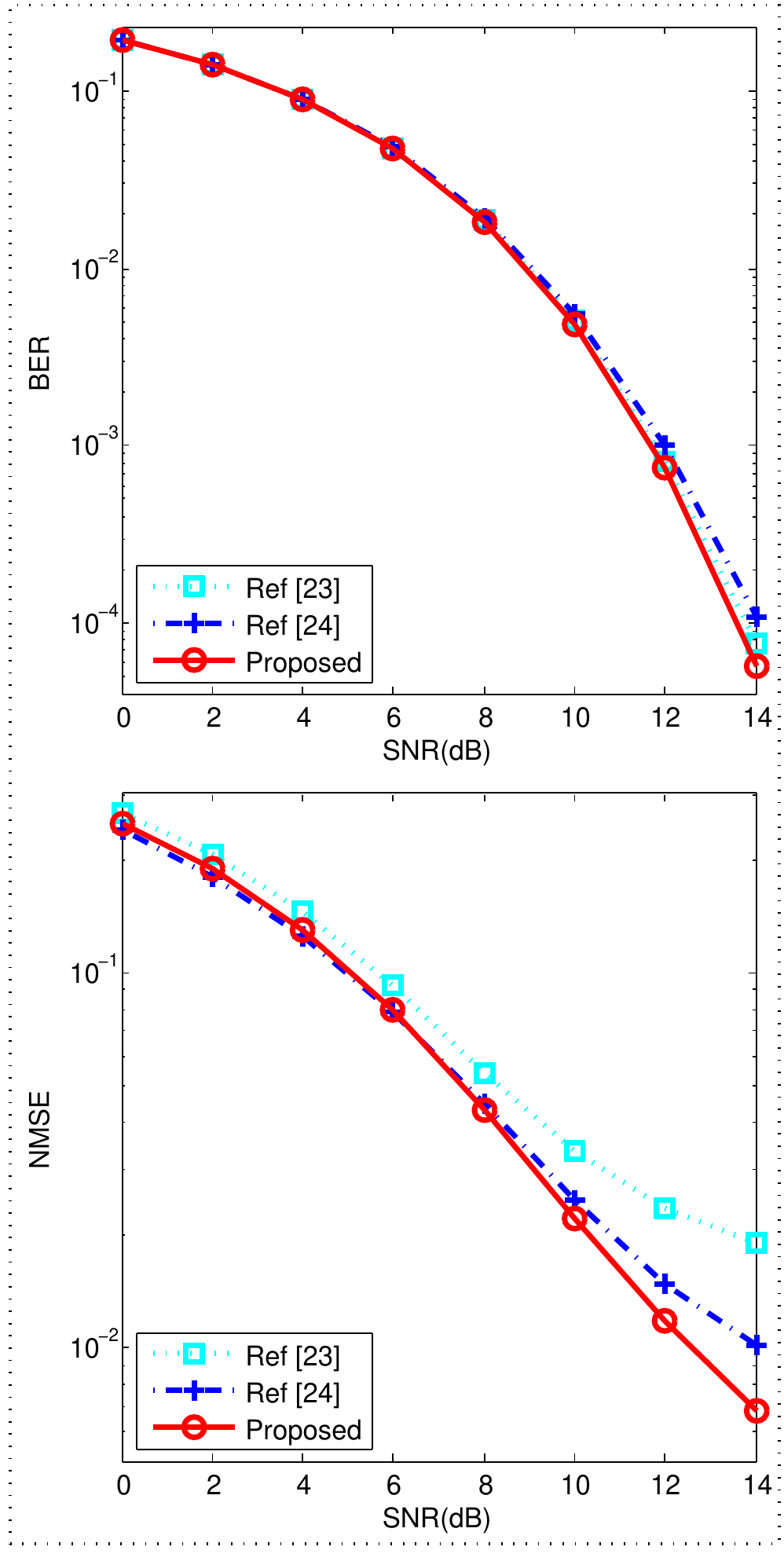}}
    \subfigure[ $N=32$]{
    \label{figure3_support_set_mse}
    \includegraphics[width=0.63\columnwidth]{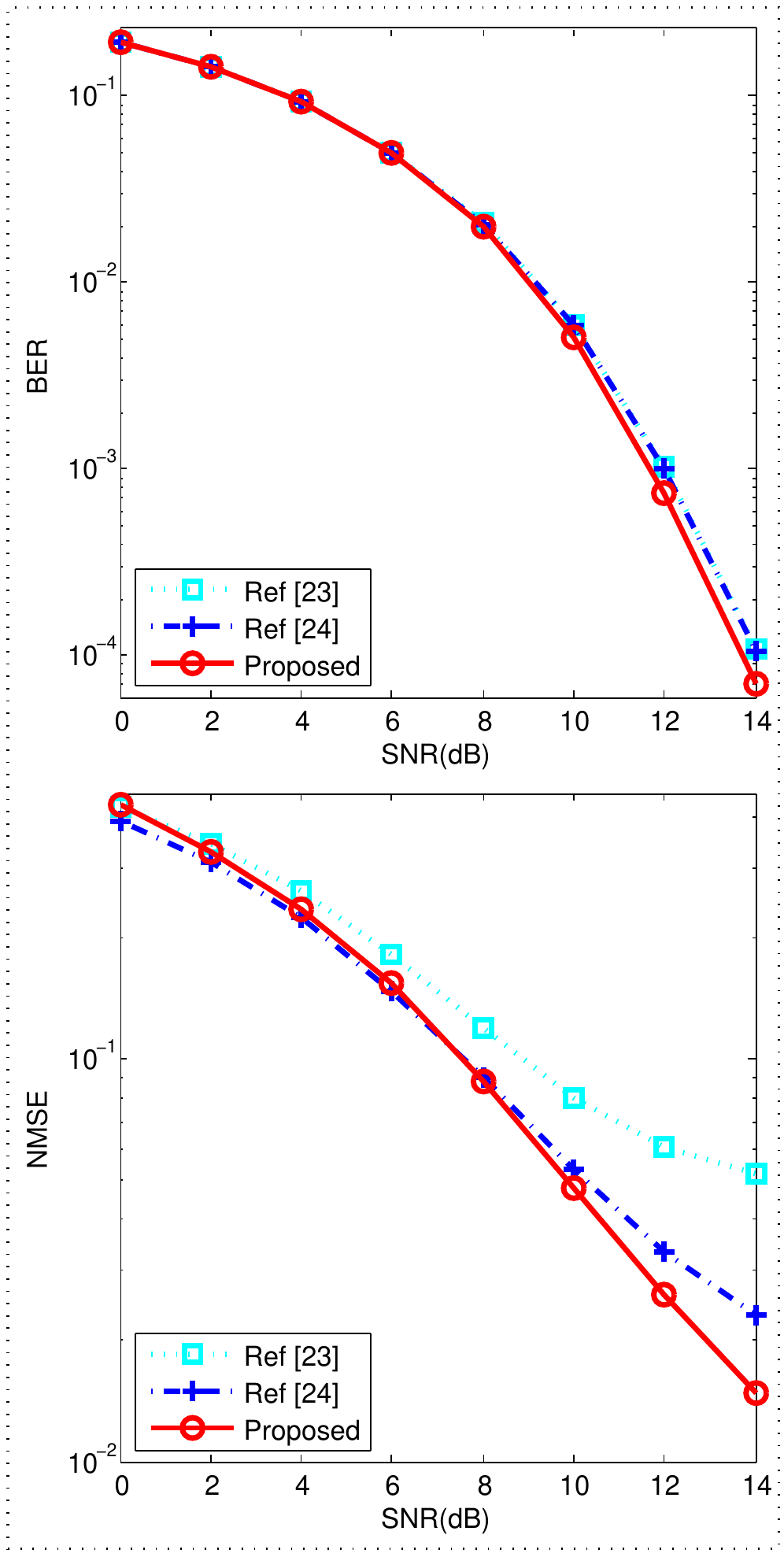}}
    \subfigure[ $N=64$]{
    \label{figure4_support_set_mse}
    \includegraphics[width=0.63\columnwidth]{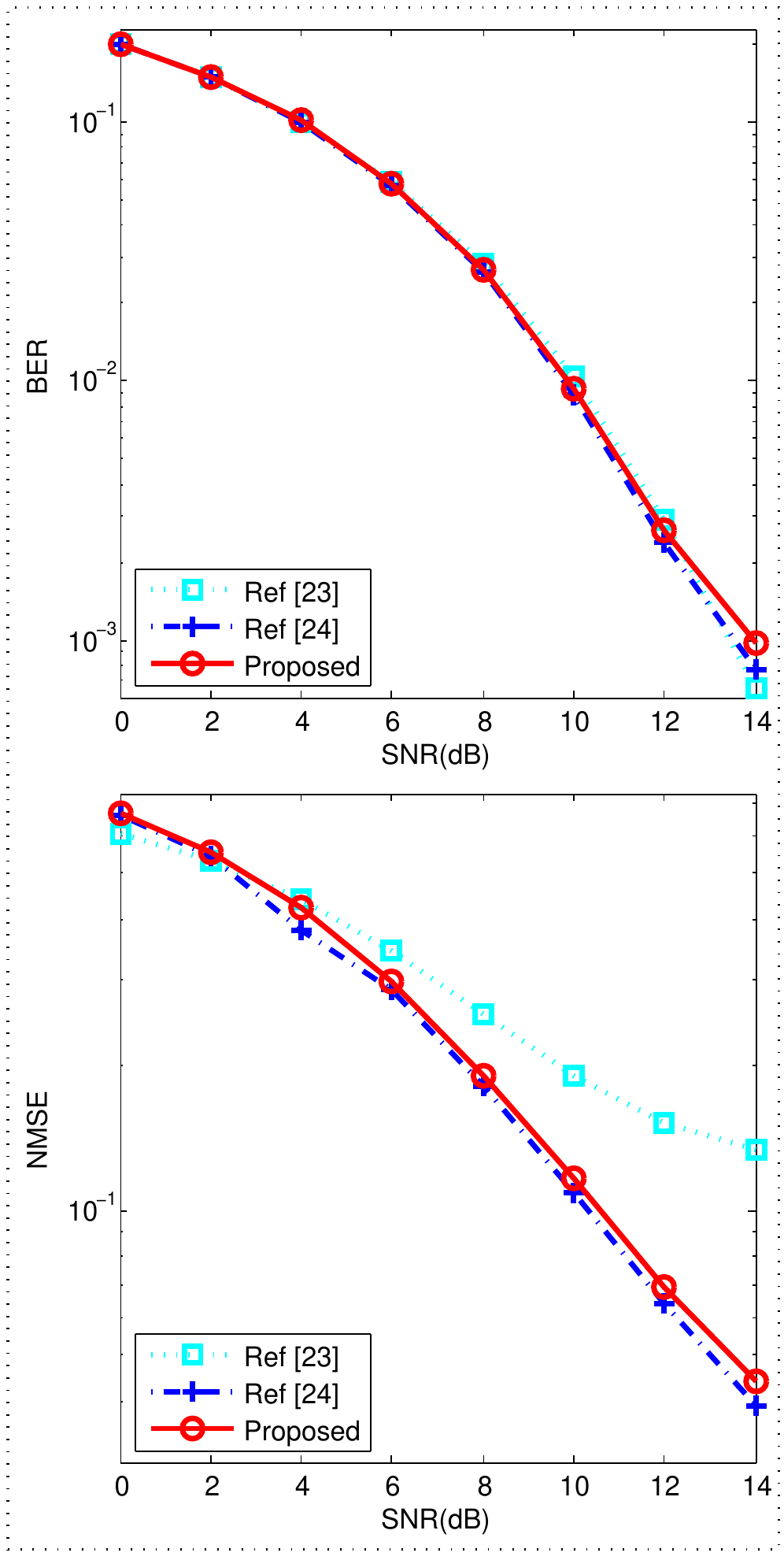}}
    \caption{BER and NMSE versus SNR, where $M{\rm{ = 512}}$, $\rho {\rm{ = }}0.20$.}
\label{fig2}
\end{figure*}

\subsection{BER AND NMSE PERFORMANCE}
\begin{figure*}[!hbtp]
\centering
    \subfigure[$\rho=0.05$]{
    \label{figure61_support_set_ber}
    \includegraphics[width=0.628\columnwidth]{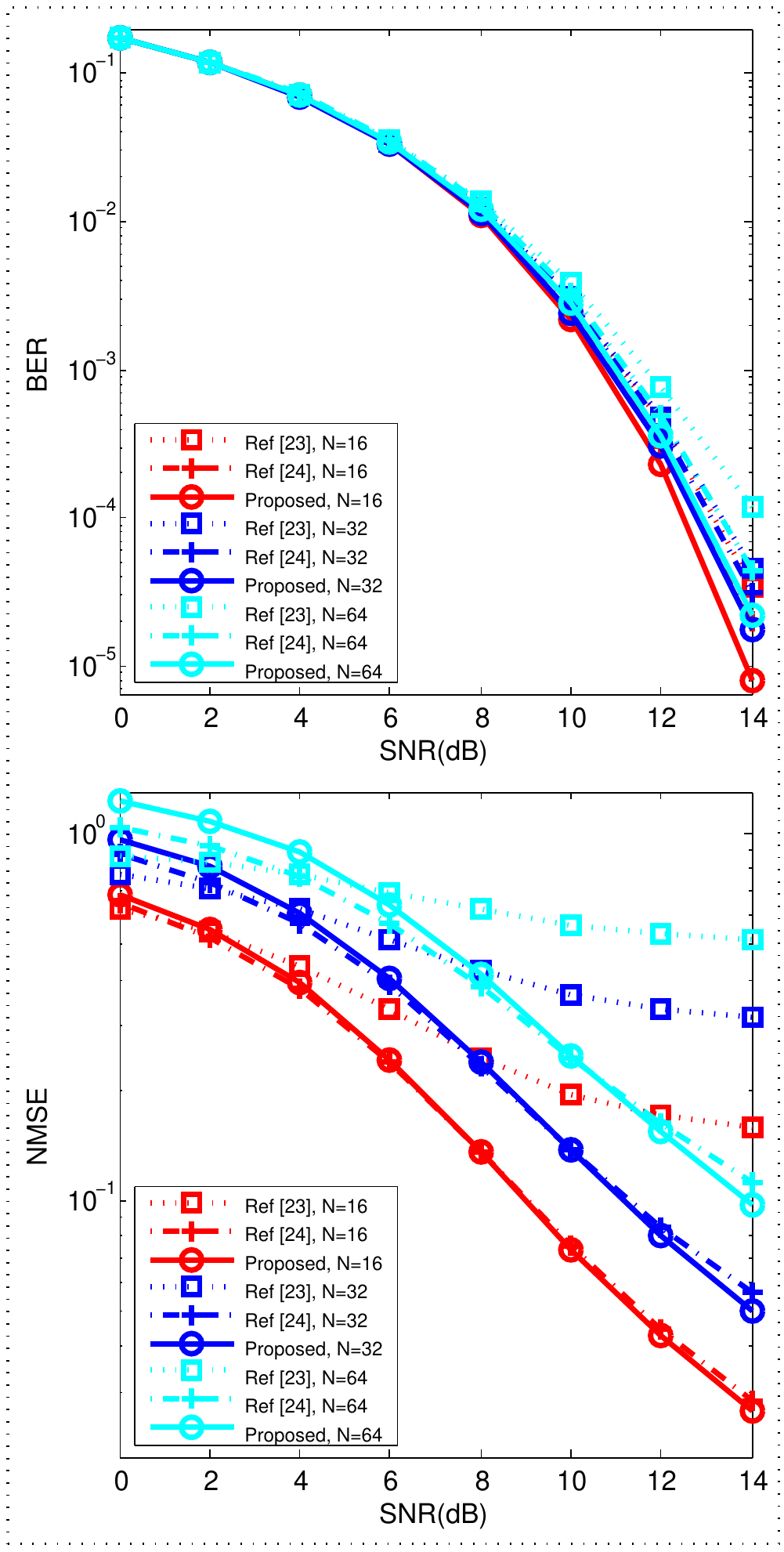}}
    \subfigure[ $\rho=0.10$]{
    \label{figure62_support_set_mse}
    \includegraphics[width=0.63\columnwidth]{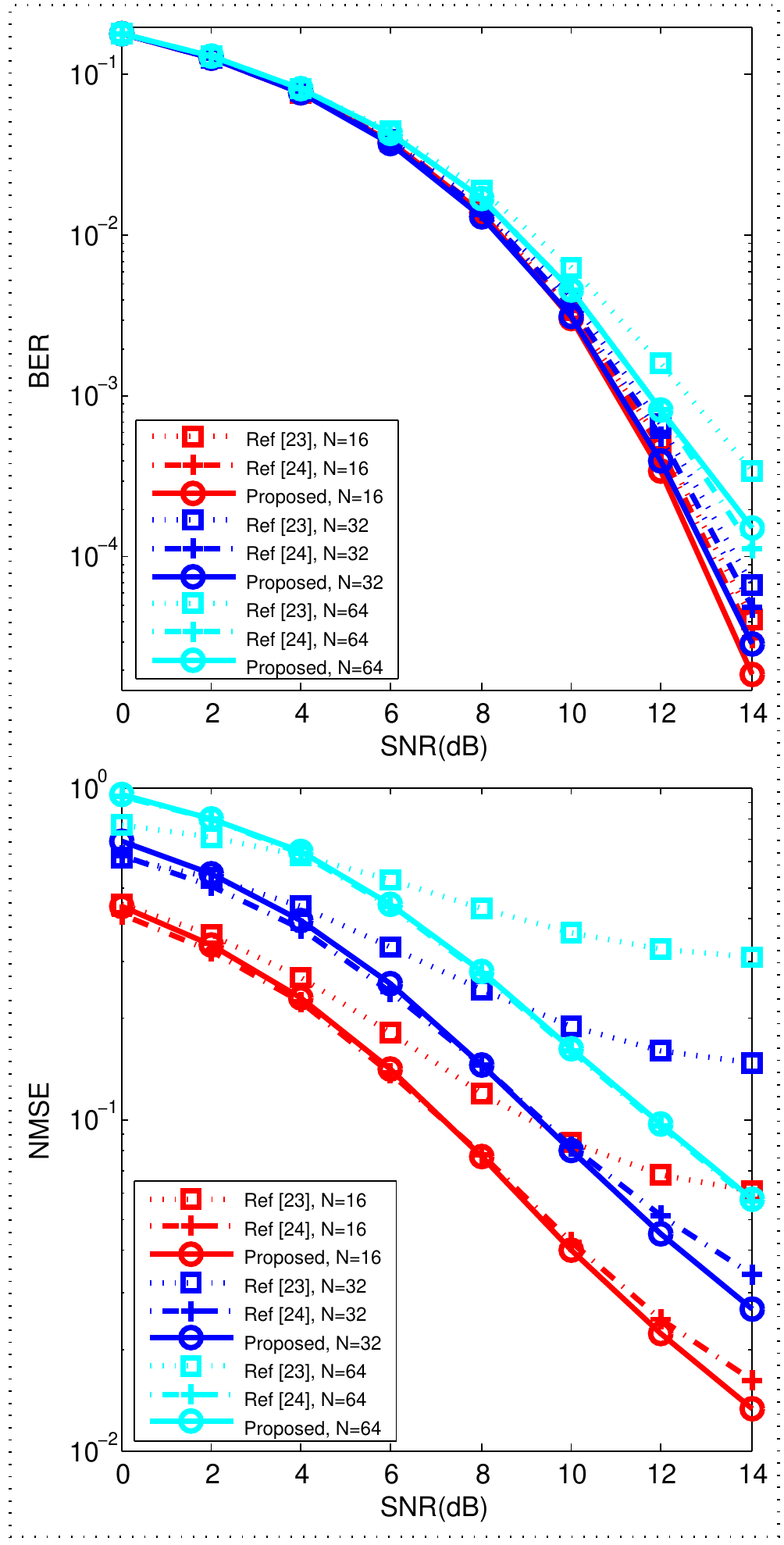}}
    \subfigure[ $\rho=0.15$]{
    \label{figure63_support_set_mse}
    \includegraphics[width=0.63\columnwidth]{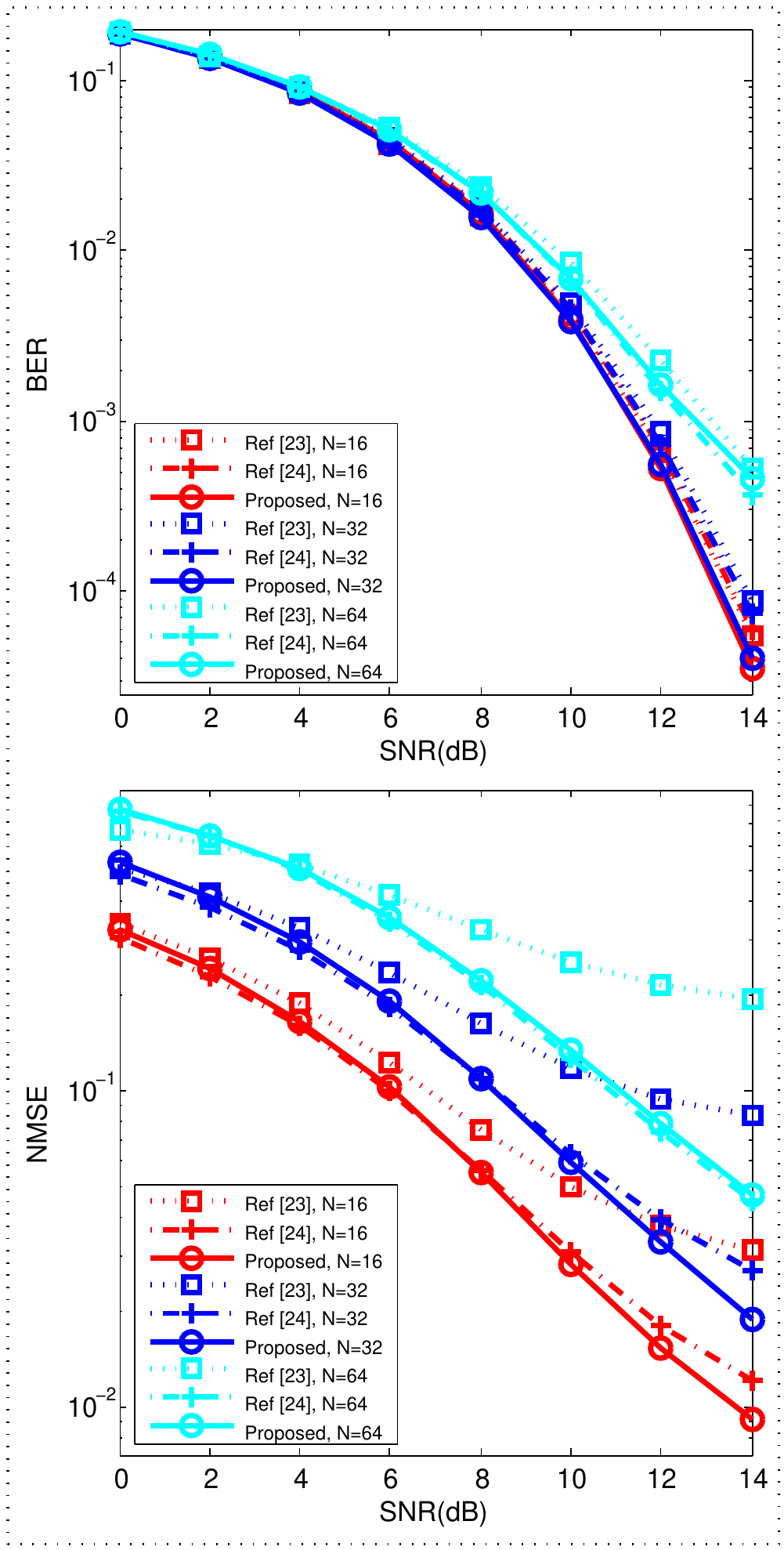}}
    \caption{BER and NMSE versus SNR, where $M{\rm{ = 512}}$, and $N$=16, 32, 64 under different values of PPC: (a)$\rho$=0.05, (b)$\rho$=0.10, and (c)$\rho$=0.15.}
\label{fig3}
\end{figure*}

In this subsection, the BER and NMSE are plotted to verify the effectiveness of the proposed ELM-based CSI feedback. We first show the performance with different values of $N$ in Fig.~\ref{fig2}. Then, the impact of PPC on BER and NMSE is given in Fig.~\ref{fig3}.

To verify the proposed method can achieve the similar performance of ``Ref\cite{a7}'', the BER and NMSE curves are presented in Fig.~\ref{fig2}, where $M{\rm{ = 512}}$, $\rho {\rm{ = }}0.20$, and different values of $N$ (i.e., $N=16$, $N=32$, and $N=64$) are considered. From Fig.~\ref{fig2}, the similar BER performances of ``Ref\cite{a16}'', ``Ref\cite{a7}'', and ``Proposed'' are observed for the same PPC $\rho {\rm{ = }}0.20$. For the NMSE performance, both ``Ref\cite{a7}'' and ``Proposed'' outperform ``Ref\cite{a16}'' due to the introducing of neural network (NN). Need to be mentioned that, the superimposed CSI feedback is a multi-task problem, i.e., both the downlink CSI and the UL-US need to be recovered at BS, and NN is particularly suitable for solving this complex problems. Besides, in terms of NMSE, ``Proposed'' is similar or slightly better than ``Ref\cite{a7}''. In detail, when $N=64$, the ``Proposed'' and ``Ref\cite{a7}'' have the similar NMSE. Yet the NMSE of ``Proposed'' is better than that of ``Ref\cite{a7}'' as $N=16$ and $N=32$. A relatively small $N$, e.g., $N \leq 32$, is easier to present this ELM-based network's advantages of NMSE. The ``Proposed'' obtains slightly better NMSE than that of ``Ref\cite{a7}''. One of the possible reasons is that the complex parameter tuning for DL-based CSI feedback in Ref\cite{a7} results in the difficulty in learning optimal network parameters. The other possible reason is that the testing SNR is not the training SNR for ``Ref\cite{a7}'', and thus degrades the NMSE performance. This also reflects that the proposed ELM-based network has a good generalization against the varying SNR. Although the BER and NMSE of ``Ref\cite{a7}'' are not improved significantly, the proposed ELM-based CSI feedback embodies obvious advantages during offline training and online running (which will be expatiated in IV-C).

In Fig.~\ref{fig3}, we validate the BER and NMSE's robustness of the proposed method against the impact of PPC, where different values of $N$ (i.e., $N=16$, $N=32$, and $N=64$) and different values of PPC (i.e., $\rho=0.05$, $\rho=0.10$, and $\rho=0.15$) are considered, respectively. From Fig.~\ref{fig3}, the BER of ``Proposed'' is slightly better than those of ``Ref\cite{a16}'' and ``Ref\cite{a7}'' for each PPC in high SNR regime (e.g., $SNR \geq 12$dB) when $N=16$ and $N=32$. The ``Ref\cite{a7}'' faces with complex parameter tuning and generalization problem, resulting in its BER performance slightly inferior to that of the proposed method. For NMSE, in low SNR regime (e.g., $SNR \leq 2$dB), the ``Ref\cite{a16}'' and ``Ref\cite{a7}'' are slightly better than ``Proposed'' for a relatively small PPC (e.g.. $\rho \leq 0.10$). The possible reasons for this are that the insufficient spread spectrum gain (defined as $M/N$) and the least-square solution (ELM transforms the learning training problem into solving the least-square norm problem of output weight matrix) result in the difficulty of resisting the noise in low SNR regime. Therefore, the proposed ELM-based network cannot work well in relatively low SNR (e.g., $SNR \leq 2$dB) with a relatively large $N$ (e.g., $N=64$) and a relatively small PPC (e.g., $\rho \leq 0.10$). In high SNR regime (e.g., $SNR \geq 12$dB), the ``Proposed'' obtains the best NMSE performance, and a larger PPC obtains greater improvements. As a whole, the proposed ELM-based superimposed CSI feedback avoids complex parameter tuning and possesses good generalization ability. In addition, the improvement in offline training and online running also makes the proposed method attractive.

To sum up, according to Fig.~\ref{fig2} and Fig.~\ref{fig3}, the proposed ELM-based CSI feedback shows similar BER and NMSE as ``Ref\cite{a7}''. In proposed method, the performances of BER and NMSE are robust against the impact of PPC. With similar BER and NMSE of ``Ref\cite{a7}'', in following subsection, we will elaborate the advantages of ``Proposed'', e.g., less training parameters, storage space and training time, etc.

\begin{table*}[!t]

\renewcommand{\arraystretch}{1.2}

\caption{OVERHEAD COMPARISON}
\label{table_III}
\centering
\begin{tabular}{|c|c|c|c|c|c|c|}
\hline
                             & \multicolumn{3}{c|}{Ref\cite{a7}}      & \multicolumn{3}{c|}{Proposed}     \\ \hline
\textit{N}                   & 16         & 32         & 64         & 16        & 32        & 64        \\ \hline
Number of training parameter & 33,606,208 & 33,705,088 & 34,099,456 & 4,198,400 & 4,210,688 & 4,259,840 \\ \hline
Storage space                & 128.198MB  & 128.575MB  & 130.079MB  & 40.031MB  & 40.125MB  & 40.500MB  \\ \hline
Training time                & \multicolumn{3}{c|}{$\geq$ 70 minutes}      & \multicolumn{3}{c|}{$\leq$ 12 minutes}   \\ \hline
Online running time          & \multicolumn{3}{c|}{$\geq$ 72 s}            & \multicolumn{3}{c|}{ $\leq$ 53 s}          \\ \hline
\end{tabular}
\end{table*}

\subsection{overhead COMPARISON}
As can be seen in IV-B, the BER and NMSE of ``Proposed'' are similar as those of ``Ref\cite{a7}''. In this subsection, we will show the overhead comparisons between ``Proposed'' and ``Ref\cite{a7}''. Their overheads are presented in TABLE~\ref{table_III}, where the number of training parameter, storage space, training time and online running time, are compared and details are described as follows.

\textbf{Number of training parameters}: Since only output weights need to be trained for each subnet, training parameters of ``Proposed'' can be computed according to $2 \times (M \times 8M) + 2\times(N \times 8N) = 16{M^2} + 16{N^2}$, where the  training parameters of CSI-ELM$i$ and DET-ELM$i$, $i=1,2$, are $8{M^2}$ and $8{N^2}$, respectively. In contrast, the ``Ref\cite{a7}'' has to train input weights, output weights, hidden biases and output biases, resulting in the training parameters are significantly increased. Furthermore, the DL-based superimposed CSI feedback in \cite{a7} employs real-valued network. The input neurons, hidden neurons, and output neurons of each subnet of ``Ref\cite{a7}'' are twice as many as those of ``Proposed'', respectively. From \cite{a7}, the number of training parameter for ``Ref\cite{a7}'' is $2 \times ( (16M \times 2M) + (2M \times 16M) + (16N \times 2N) + (2N \times 16N) + 16M + 2M + 16N + 2N )= 128{M^2} + 36M + 128{N^2} + 36N$, which is obviously larger than that of ``Proposed'' (i.e., $16{M^2} + 16{N^2}$). For the examples of $N=16$, $N=32$, and $N=64$, TABLE~\ref{table_III} also verifies the training parameters of ``Proposed'' are less than those of ``Ref\cite{a7}''. Therefore, compared with DL-based superimposed CSI feedback in \cite{a7}, the proposed ELM-based superimposed CSI feedback significantly reduces the number of training parameters.

\textbf{Storage space}:  As for the storage space, the float32 data type is considered, i.e., a data is stored by using 4 bytes. In ``Ref\cite{a7}'', the memories for input weights, output weights, hidden biases and output biases are required. Thus the storage space is $(128{M^2} + 36M + 128{N^2} + 36N)\times 4 = 512({M^2} +{N^2})+144(M+N)$ bytes. For ``Proposed'', the output weights need $(16{M^2} + 16{N^2}) \times 2 \times 4 = 128({M^2} + {N^2})$ bytes, where we product 2 for the complex-valued entry of output weights. In addition, both the input weights and hidden biases of each subnet are loaded from a same matrix $\mathbf{W}$, leading to the storage space is $\mathrm{max}\{(8M^2 \times 4), (8N^2 \times 4) \}$ bytes. Usually, $\mathrm{max}\{(8M^2 \times 4), (8N^2 \times 4)\}=32M^2$ due to the main consideration of user service (the length of UL-US is larger than the lenght of downlink CSI, i.e., $M>N$). Then, the storage space of ``Proposed'' is $128({M^2} + {N^2})+32{M^2} = 160{M^2} +128{N^2} $ bytes. Relative to ``Ref\cite{a7}'', the ``Proposed'' saves the storage space $(512({M^2} +{N^2})+144(M+N))-(160{M^2} +128{N^2}) = 352{M^2} +384{N^2} + 144(M+N) $ bytes. From TABLE~\ref{table_III}, the significant memory saving from the ``Proposed'' is observed compared to ``Ref\cite{a7}'', e.g., for $N = 32$, ``Ref\cite{a7}'' needs the storage space 128.575MB while the storage space for ``Proposed'' is only 40.125MB.

\textbf{Training time}: Since the network of ``Ref\cite{a7}'' is DL-based network, thus we carry out its training on a server with NVIDIA TITAN RTX GPU and Intel Xeon(R) E5-2620 CPU 2.1GHz$\times$16. Unlike the DL-based network of ``Ref\cite{a7}'', the proposed ELM-based network is a complex-valued and feed-forward network, yielding its training can be easily performed by using Matlab's matrix operation on a personal computer (PC). Thus, we train the network of ``Proposed'' on a PC with CPU i5-4210U (1.7GHz$\times$4). From the comparison of training parameters, the training parameters of ``Proposed'' are far fewer than those of ``Ref\cite{a7}'', leading to a shorter training time. In TABLE~\ref{table_III}, the maximum time consumption of ``Proposed'' and the minimum time consumption of ``Ref\cite{a7}'' are considered among the cases where $N=16$, $N=32$, and $N=64$. Even without GPU's acceleration, to capture similar BER and NMSE, the training time of ``Proposed'' is less than 12 minutes, while the ``Ref\cite{a7}'' consumes more than 70 minutes. Overall, compared with ``Ref\cite{a7}'', the proposed method significantly reduces the training time.

\textbf{Online running time}: Because the same expert knowledge is adopted by ``Proposed'' and ``Ref\cite{a7}'', the main difference of online running time is reflected in four subnets. Thus, we observe and analyse the online running time according to the four subnets.  For a fair comparison, $10^4$ online-running experiments are conducted for ``Proposed'' and ``Ref\cite{a7}'' on the same PC (with CPU i5-4210U) by using Matlab software, respectively. That is, the DL-based method in ``Ref\cite{a7}'' is also run on Matlab software by importing its trained network parameters and architecture. During the experiments, only running time is considered, i.e., the time for generating the data set is not included. From TABLE~\ref{table_III}, the time consumption of ``Ref\cite{a7}'' is much longer than that of ``Proposed''. The analysis can also derive the same conclusion. For ``Proposed'', each DET-ELM$i$ has $2 \times (8M \times M) + 4\times(M \times 8M) =48M^2$ real multiplications, and each CSI-ELM$i$ has $2 \times (8N \times N) + 4\times(N \times 8N) =48N^2$ real multiplications. Here, the input weights and output weights are real-valued matrix and complex-valued matrix, respectively. Relative to complex-valued input, the input weights and output weights respectively produce 2 times and 4 times real multiplications. In addition, the real additions of DET-ELM$i$ and CSI-ELM$i$ are $2 \times [8M \times (M - 1) + 8M] + 2 \times [M \times (8M - 1 + 8M)] = 48M^2 - 2M$ and $2 \times [8N \times (N - 1) + 8N] + 2 \times [N \times (8N - 1 + 8N)] = 48N^2 - 2N$, respectively. As a whole, the ``Proposed'' has $2 \times (48M^2 + 48N^2) = 96(M^2 + N^2)$ real multiplications and $2 \times [(48M^2 - 2M) + (48N^2 - 2N)] = 96(M^2 + N^2) - 4(M + N)$ real additions. Unlike the ``Proposed'', the DL-based superimposed CSI feedback in \cite{a7} employs real-valued network. The input neurons, hidden neurons, and output neurons of each subnet of ``Ref\cite{a7}'' are twice as many as those of ``Proposed'', respectively. Thus, the number of multiplications and additions can be computed as $2 \times \{[(16M \times 2M) + (2M \times 16M)] + [(16N \times 2N) + (2N \times 16N)]\} = 128(M^2 + N^2)$ and $2 \times \{[16M \times (2M - 1) + 16M + 2M \times (16M - 1) + 2M] + [16N \times (2N - 1) + 16N + 2N \times (16N - 1) + 2N]\} = 128(M^2 + N^2)$, respectively. Compared with ``Ref\cite{a7}'', the ``Proposed'' saves $128(M^2 + N^2) - 96(M^2 + N^2) = 32(M^2 + N^2)$ real multiplications and $128(M^2 + N^2) - [96(M^2 + N^2) - 4(M + N)] = 32(M^2 + N^2) + 4(M + N)$ real additions, and thus reduces the online running time.

From the overhead comparison in this subsection, the proposed ELM-based CSI feedback embodies many advantages. Concretely, with similar BER and NMSE, the ``Proposed'' has less training parameters, storage space, offline training time and online running time than ``Ref\cite{a7}''.

\section{CONCLUSION}
The DL-based superimposed CSI feedback is still facing many challenges, such as the complexity of parameter tuning, huge number of training parameters, long offline-training and online-running time, etc. To remedy these defects, the ELM-based superimposed CSI feedback has been investigated in this paper.
By employing the simplified versions of ELM-based subnets, the proposed method brings little change to the neural network structure of the original DL-based network but significantly reduces training parameters and offline-training time. More importantly, without loss of BER and NMSE performance, the proposed method requires less storage space and online-running time than those of original DL-based superimposed CSI feedback. Other approaches such as combining compression techniques and ELM could also be explored in future work.


\begin{IEEEbiography}[{\includegraphics[width=1in,height=1.25in,clip,keepaspectratio]{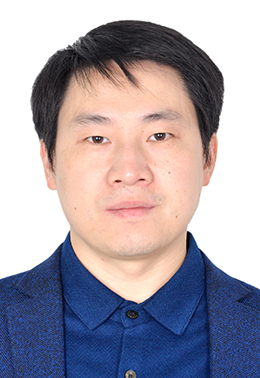}}]{Chaojin Qing} (M'15) received the B.S. degree in communication engineering from Chengdu University of Information Technology, Chengdu, China, in 2001, the M.S. and Ph.D. degrees in communications and information systems from the University of Electronic Science and Technology of China, Chengdu, China, in 2006 and 2011, respectively. From November 2015 to December 2016, he was a Visiting Scholar with Broadband Communication Research Group (BBCR) of the University of Waterloo, Waterloo, ON, Canada.

From 2001 to 2004, he was a teacher with the Communications Engineering Teaching and Research Office, Chengdu University of Information Technology, Chengdu, China. Since 2011, he has been an Assistant Professor with the School of Electrical Engineering and Electronic Information, Xihua University, Chengdu, China. He is the author of more than 40 papers and more than 20 chinese inventions. His research interests include detection and estimation, massive MIMO systems, and deep learning in physical layer of wireless communications.

\end{IEEEbiography}

\begin{IEEEbiography}[{\includegraphics[width=1in,height=1.25in,clip,keepaspectratio]{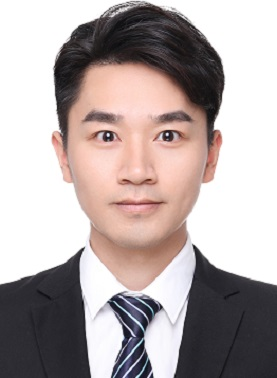}}]{Bin Cai}
received the B. S. degree from the School of Electrical Engineering and Electronic Information, Xihua University, Chengdu, China, in 2018, where he is currently  pursuing the M. S. degree under the supervision of Prof. Qing. His research interests include detection and estimation, channel state information feedback, and deep learning in physical layer of wireless communications.
\end{IEEEbiography}

\begin{IEEEbiography}[{\includegraphics[width=1in,height=1.25in,clip,keepaspectratio]{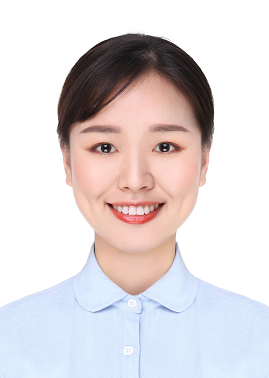}}]{Qingyao Yang} received the B. S. degree from the School of Electrical Engineering and Electrical Information, Xihua University, Chengdu, China, in 2017, where she is currently pursuing the M. S. degree under the supervision of Prof. Qing. Her research interests include compressed sensing, channel state information feedback, signal processing in wireless communications, and deep learning in physical layer of wireless communications.
\end{IEEEbiography}

\begin{IEEEbiography}[{\includegraphics[width=1in,height=1.25in,clip,keepaspectratio]{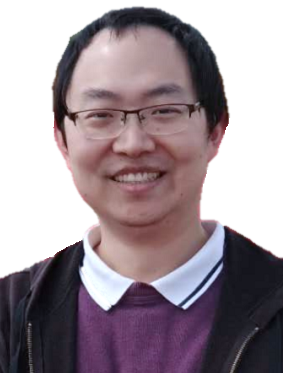}}]{Jiafan Wang} (S'15) received his B.S. degree and M.S. degree in Electrical Engineering from University of Electronic Science and Technology of China in 2006 and 2009, respectively. He accomplished the Ph.D.  degree in Computer Engineering at Texas A$\&$M University, College Station, TX, USA in 2017.

His major field of study is smart integrated circuit design, which includes multi-dimensional non-deterministic gate implementation with systematic optimization framework, self-training Analog/Digital Mixed-System for circuit feature calibration, and configurable locking mechanism against Analog IP piracy. He is currently working in Synopsys Inc. to develop the world's leading silicon chip design software in electronic design automation (EDA) industry.
\end{IEEEbiography}

\begin{IEEEbiography}[{\includegraphics[width=1in,height=1.25in,clip,keepaspectratio]{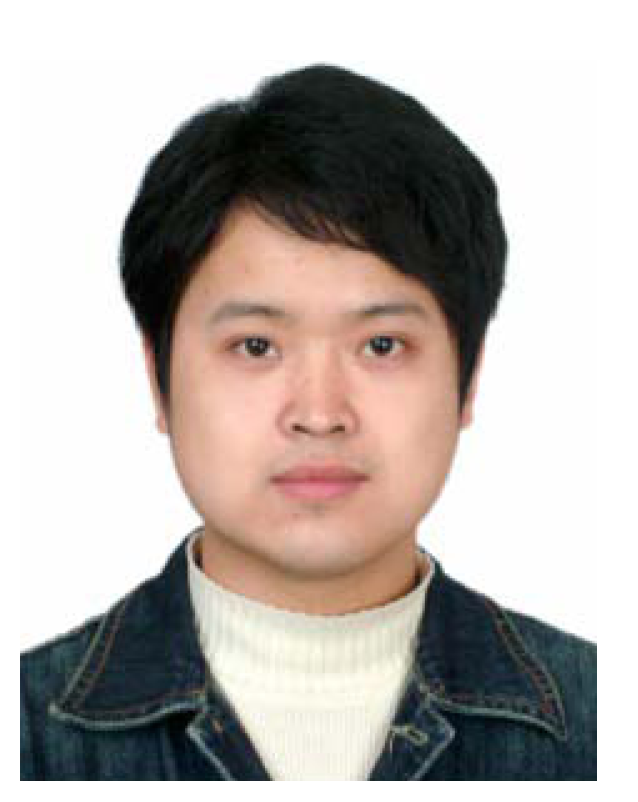}}]{Chuan Huang} (S'09--M'13) received the B.S. degree in math and the M.S. degree in communications engineering from the University of Electronic Science and Technology of China, Chengdu, and the Ph.D. degree in electrical engineering from Texas A$\&$M University, College Station, TX, USA, in 2012. From 2012 to 2013, he was with Arizona State University, Tempe, AZ, USA, as a Post-doctoral Research Fellow, and then promoted to Assistant Research Professor from 2013 to 2014. He was also a Visiting Scholar with the National University of Singapore and a Research Associate with Princeton University.

He is currently with the National Key Laboratory of Science and Technology on Communications, University of Electronic Science and Technology of China. His current research interests include energy harvesting communications, multicast traffic scheduling, full-duplex communications, and signal processing in wireless communications. He has served as a TPC member for many IEEE conferences.
\end{IEEEbiography}

\EOD

\end{document}